\documentstyle[prb,aps,epsfig,twocolumn]{revtex}

\draft
\title{\bf Approach to the metal-insulator transition in La$_{1-x}$Ca$_{x}$MnO$_3$ $(0\leq x \leq 0.2)$:  magnetic inhomogeneity and spin-wave anomaly}
\author{ G. Biotteau$^1$, M.Hennion$^1$, F. Moussa$^1$, 
J. Rodr\'{\i}guez-Carvajal$^1$, L. Pinsard$^{2}$, A. Revcolevschi$^2$, Y. M. Mukovskii$^3$ and D. Shulyatev$^3$}

\vspace{0.5cm}

\address{$^1$Laboratoire L\'eon Brillouin, CEA-CNRS, CE Saclay, 
91191 Gif sur Yvette Cedex, France}
\address{$^2$Laboratoire de Chimie des Solides, 
Universit\'e Paris-Sud, 91405 Orsay Cedex, France}
\address{$^3$Moscou State Steel and Alloys Institute, Leninskii prospect 4, Moscou 117936 Russia} 
\date{\today, submitted to Phys. Rev. B}
\begin{document}
\twocolumn[\hsize\textwidth\columnwidth\hsize\csname @twocolumnfalse\endcsname
\maketitle
\begin{abstract}
We describe the evolution of the static and dynamic spin correlations of La$_{1-x}$Ca$_x$MnO$_3$, for x=0.1, 0.125 and 0.2,  where the system evolves from the canted magnetic state towards the insulating ferromagnetic state, approaching the metallic transition (x=0.22). 
 In the x=0.1 sample, the observation of two spin wave branches typical of two distinct
types of magnetic coupling, and of a modulation in the elastic diffuse scattering characteristic of ferromagnetic inhomogeneities, confirms the static and dynamic inhomogeneous features previously observed at x$<$0.1. 
The anisotropic 
q-dependence of the intensity of the low-energy spin wave suggests a bidimensionnal character for the static inhomogeneities. At x=0.125, which corresponds to the occurence of a ferromagnetic and insulating state, the two spin wave branches reduce to a single one, but anisotropic. 
At this concentration, an anomaly appears at {\bf q$_0$}=(1.25,1.25,0), that could be related to an underlying periodicity, as arising from (1.5,1.5,0) superstructures.
 At  x=0.2, the spin-wave branch is isotropic. In addition to the anomaly observed at q$_0$, extra magnetic excitations are observed at larger q, forming an optical branch. The two dispersion curves suggest an anti-crossing behavior at some {\bf q$_0$'} value, which could be explained by a folding
 due to an underlying perodicity involving
 four cubic lattice spacings. 
\end{abstract} 
 ] 
 
\centerline{\small \bf I. INTRODUCTION}
\vspace{0.5cm}

The doped rare-earth manganites have now been studied for several decades\cite{Wollan,Goodenough}. They have raised up a renewed
interest because of their colossal magnetoresistance observed beyond a critical value of doping rate. However, the physics driving these properties are still unclear. At low doping, the nature of the magnetic ground state is still a well-debated issue. From a theoretical point of view, depending on parameter values, the foundamental state can be either
 an homogeneous canted antiferromagnetic state\cite{de Gennes,Arovas} or an					 inhomogeneous state
 with a phase separation between hole-rich $(Mn^{4+})$ ferromagnetic regions and hole poor $(Mn^{3+})$ antiferromagnetic
 domains\cite{Arovas,Khomskii,Kagan,Nagaev,Dagotto,Yunoki}. At higher concentrations, the existence of a ferromagnetic and insulating phase is also very intringuing. It stresses out that the double-exchange (DE) coupling 
alone, which predicts a ferromagnetic and metallic state is unsufficient to explain the magnetic properties\cite{Millis1,Roder}. 
 Around a doping rate of x=0.125, several works on the Sr-doped compounds have reported anomalous structural and magnetic properties at low temperature\cite{kawano2,pinsard}, interpreted in terms of a new orbital ordering\cite{Endoh} in these systems. A comparison with Ca-substitution is needed  for a deeper understanding of these effects.

Neutron scattering is a very powerful technique to describe, at an atomic scale, the evolution of a magnetic system from an insulating state to a metallic one.
 In previous papers devoted to the $x = 0, x = 0.05$ and $x = 0.08$ of Ca concentration\cite{Moussa1,Hennion1,Hennion3,Moussa2}  
  several features have been established. i) Bragg peaks, at low temperature indicate a mean canted antiferromagnetic state. 
 ii) Diffuse scattering is observed, characteristic of ferromagnetic inhomogeneities of characteristic size, distributed in a quasi-liquid order.
 These inhomogeneities have been attributed to hole-rich regions, embedded in an hole-poor medium. iii) The spin dynamics, where a high-energy and a low-energy spin-wave branches are observed, reflects both the mean antiferromagnetic canted state 
 and the inhomogeneous features.
The high energy spin wave, can be described
 by two superexchange (SE) couplings, as in pure LaMnO$_{3}$, characteristic of the A-layered antiferromagnetic structure.
 The low energy spin-wave branch is characteristic of an isotropic ferromagnetic coupling. The q-dependence of its susceptibility
 reveals a tight connection with
  the ferromagnetic inhomogeneities. This new ferromagnetic coupling has been suggested to be related to double-exchange (DE).
 So, 
 the $Ca$-doped compounds, at low values of doping, exhibit both {\it homogeneous} and {\it inhomogeneous} 
 magnetic features. The same observations have been found in a Sr-doped compound\cite{Hennion2}, showing the general character of these observations in low-doped manganites. 
 
 In this paper we present a new neutron scattering study at three concentrations, 0.1, 0.125, and 0.2, allowing a general survey of the concentration range where the system, still an insulator, approaches the metallic state. The phase diagram, established in the 0$\le$x$\le$0.2 range is pecularly studied  in the 0.125$\le$x$\le$0.2 range, where, below the ferromagnetic transition temperature, a structural transition
 appears. The spin dynamics is determined along the two relevant q-directions for the A-type structure, [110] and [001]. In the $x = 0.10$ compound which exhibits many 
  features similar to the 0$<$x$<$0.1 compounds, we observe a strong anisotropy in the q-dependence of the low-energy spin wave intensity, that we relate to a two-dimensional character for the static ferromagnetic clusters. At $x = 0.125$, {\it an abrupt change} occurs. We no more observe a modulation in the diffuse scattering, suggesting a percolation of the ferromagnetic clusters. Moreover, the spin dynamics reduces to a single spin wave branch, anisotropic, with an anomaly at some $q_0$=(1.25,1.25,0).
At $x=0.2$, the spin wave dispersion is isotropic. Whereas the most intense excitations allow to define the dispersion curve of a usual ferromagnet with a small anomaly at $q_0$ as for $x=0.125$, additionnal magnetic excitations are observed at larger q, forming an optical branch. They reveal a more complex magnetic state, suggesting a high sensitivity of the magnons to an underlying structural periodicity.

The paper is organized as follows. Experimental details are given below,
 in this introduction. The structural and magnetic phase diagram for the 0$\le$x$\le$0.2 concentration range is presented
   in section II. Diffuse scattering experiments obtained for x=0.05, 0.08, 0.1 and 0.125 are reported in section III. The spin dynamics observed for x=0.1,0.125 and 0.2 is described in section IV  and, finally,
    section V is devoted to a discussion and conclusion.

{

\vspace{1cm}

Single crystal of La$_{0.9}$Ca$_{0.1}$MnO$_3$ was grown by a floating zone 
method associated with an image furnace at the Laboratoire de Physico-Chimie de l'\'etat 
Solide in Orsay, France, with a volume of about 0.4 cm$^3$. Samples with higher Ca concentrations, were grown by the same method at the MISIS Institute of Moscou, with similar volumes. Preliminary results are also reported for x=0.17. The mosaicity of 
these samples is small: 0.6$^{\circ}$, 0.5$^{\circ}$ and 0.5$^{\circ}$ for the 
$x = 0.1$, $0.125$, $0.17$, $0.2$ doped compounds, respectively. The Ca concentration has been checked by comparing the lattice parameters with the previous determination of Matsumoto\cite{matsumoto}, and, for x = 0.125  
and x = 0.17, by performing surface microscopy. A good agreement is also found with the phase diagram of Cheong, published in Ref. 21, except for the sample with nominal concentration $x = 0.2$, which would  rather correspond to $x = 0.19$. Nevertheless, in the following we keep the nominal concentration x=0.2.
  All our samples are 
twinned in three space directions, but for $x = 0.1$ and $0.125$ and $x = 0.17$, the 
orthorhombicity is large enough to allow the resolution of the Bragg peaks corresponding to different domains.

Neutron scattering experiments were performed on several triple axis 
spectrometers installed either on thermal or cold neutrons source at the  Orph\'{e}e reactor of Laboratoire L\'{e}on Brillouin (1T, 4F1, 4F2, G43)
and at the reactor of Institut Laue Langevin (IN14). Elastic spectra have been 
obtained using a wave vector of $k_i = k_f = 1.55  \AA^{-1}$ associated with a 
Berylium filter and tight 10'-10' collimations. Energy spectra 
were measured using various fixed outcoming wave vectors $k_f$ 
(from 2.662 to 1.05 $\AA^{-1}$) and pyrolytic graphite or
berylium as filter. All the samples were oriented with a horizontal
(001,110) scattering plane, defined in {\it Pbnm} symmetry ($c/\sqrt{2}<a<b$) except for the $x=0.2$ single crystal which
was studied in all q directions.

Macroscopic magnetization measurements were performed at Laboratoire de Physico-Chimie 
de l'\'etat Solide in Orsay and at the SPEC (CEA-Saclay) with a SQUID magnetometer, for fields up to 5T.
\vspace{0.5cm}

\centerline{\small \bf II. STRUCTURAL AND MAGNETIC}
\centerline{\small \bf PHASE DIAGRAM}
\vspace{0.5cm}

Structural and magnetic transition temperatures have been determined from Bragg peak 
intensities using elastic neutron scattering and are reported in Fig. 1. \\

\noindent
{\it 1) Structural transition}\\

In Fig. 1, the solid line through the filled squares can be identified as the Jahn-Teller 
transition temperature $T_{OO'}$. There, the system evolves from a pseudo-cubic phase 
with a dynamical Jahn-Teller effect to an orthorhombic Pbnm phase 
with a cooperative and static Jahn-Teller effect. We observe a strong decrease of the structural transition temperature from $T_{OO'}=750 K$ in pure LaMnO$_3$
\cite{juan} to $T_{OO'}=200 K$ at $x=0.2$. Actually, $T_{OO'}$ has been determined
 from the increase of the q linewidth, which locates the temperature where a single Bragg 
peak in the O phase
splits into two ones in the O' phase due to twinning. Below $x=0.2$, the experimental q-resolution allows a determination of the cell parameters in the orthorhombic phase, as shown in Fig.2-a and Fig.2-b for x = 0.1 and x = 0.125, respectively. For x = 0.2, the Bragg peaks are not resolved whatever the temperature in the O' phase located below 200K . 
At $x$$\ge$$0.125$, a further decrease of the orthorhombicity is observed below 
100 K, indicating a re-entrance for the high-temperature pseudo-cubic phase. We call T$_B$ the temperature of the inflexion point of this variation. Very interestingly, the same anomaly has been observed in Sr-doped samples
\cite{kawano2,pinsard,vasiliu,arkhipov}. In La$_{0.875}$Sr$_{0.125}$MnO$_3$, where T$_B$ is 
called T$_{O'O"}$, this variation has been shown to correspond to a rapid change in two of the three Mn-O distances\cite{pinsard}. Although the same evolution is expected for these Ca-doped samples, such a structural study is lacking. Moreover, in Sr-doped compounds, new Bragg reflexions occur below T=T$_{O'O"}$, namely at (0,0,2l+1) which are forbidden in {\it Pbnm} symmetry\cite{pinsard,kawano2} and (0,0,l+0.5)\cite{Yamada}. They have been related either to a polaron ordering\cite{Yamada} or to a new orbital ordering\cite{Endoh1}.  In the present study, 
nuclear (0,0,2l+1) peaks are actually detected at $x=0.2$ only, with a small intensity and in the whole studied temperature range (10K$\le$T$\le$300K), whereas the superstructures (0,0,l+0.5), searched for at x=0.125 and x=0.2, have not been found. 
\\
\begin{figure}
\centerline{\epsfig{file=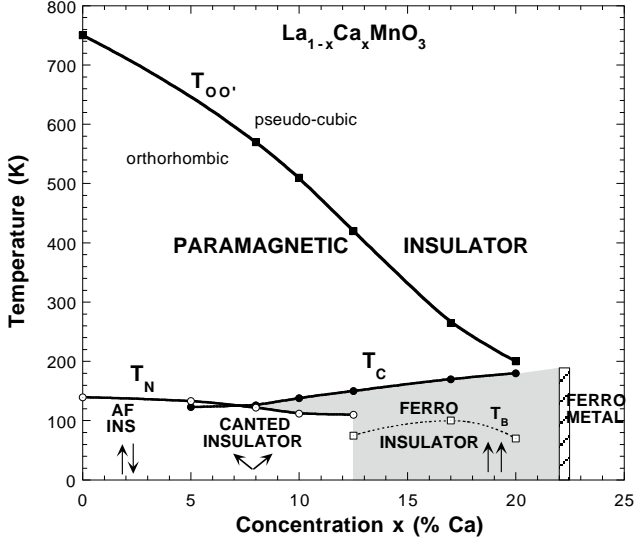,width=8.5cm}}
\vspace{0.5cm}
\caption{(T,x) phase diagram of La$_{1-x}$Ca$_{x}$MnO$_3$, determined by
neutron scattering measurements on single crystals.}
\end{figure}
\noindent

\begin{figure}
\centerline{\epsfig{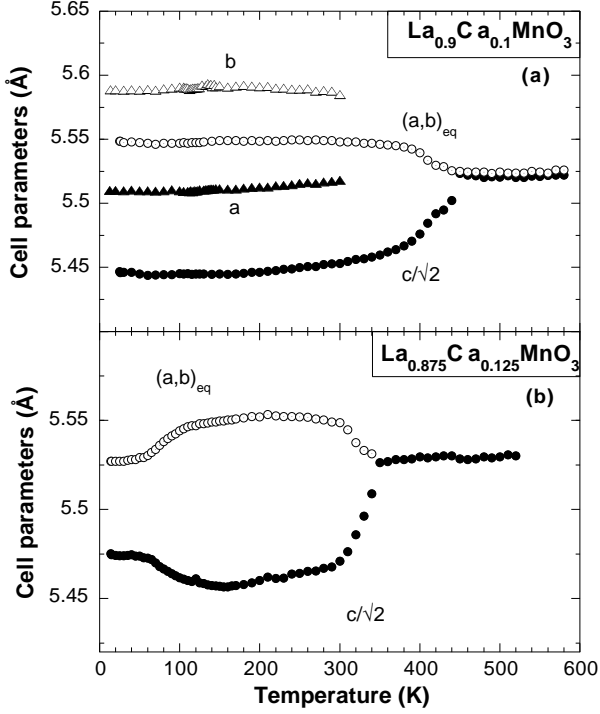}}
\vspace{0.5cm}
\caption{Cell parameters versus temperature {\bf (a)} $x=0.1$ and {\bf (b)} $x=0.125$. 
(a,b)$_{eq}$ is deduced from the (110) Bragg peak position and defined by 
$(a,b)_{eq} = \sqrt{2}/\sqrt{1/a^2+1/b^2}$.}
\end{figure}

\noindent
{\it 2) Magnetic structure}\\

As reported in Fig. 1, below $T_{OO'}$, two magnetic transitions are observed. 

{\it For $x=0.1$}, a ferromagnetic transition occurs at $T_C=138K$, and an antiferromagnetic ordering of spin components occurs at $T_{CA}=112K$.  This succession of phases has been predicted by de Gennes\cite{de Gennes}, for some value of the parameters of the mean field model. Interestingly,
the direction of the ferromagnetic spin component shows a complex evolution with temperature, reflected in the {\bf $\tau$}= (001), (002), (020) and 
(200) Bragg peak intensities reported in Fig. 3-a, 3-b, 3-c respectively. 
In the neutron cross-section, the geometrical factor implies that 
only spin components perpendicular to {\bf Q} can contribute to the scattering intensity.
 Therefore, between T$_C$ and T$_{CA}$, the observation of a constant intensity for
 the (020) Bragg peak (Fig 3-c) indicates that spins are aligned along the {\bf b} axis. At 
T$_{CA}$, Rietveld refinements indicate that an antiferromagnetic component
develops also along {\bf b}, keeping  the $A_y$ type structure found 
for $x<0.1$. The concomitant decrease of the (002) Bragg peak 
intensity (Fig 3-b) reflects a rotation of the ferromagnetic spin component from {\bf b} to {\bf c} 
axis, as antiferromagnetism develops. 
This effect was not observed at $x=0.08$, where T$_C$ and T$_{CA}$ are very close to each other, and the ferromagnetic spin component is along {\bf c} at all temperatures.
Therefore, at low temperature, the {\it mean} spin components deduced from Bragg peaks
 consist in an antiferromagnetic  component along the {\bf b} axis, a ferromagnetic one along 
the {\bf c} axis, and another small ferromagnetic component along the {\bf a} axis. From this 
spin configuration, we define a {\it mean} canting angle (angle between the spin direction and the 
direction {\bf b} of the antiferromagnetism) equal to 
$\theta_c= \arccos(\sin \theta \sin \phi) =61.5^\circ \pm 5^\circ$.

\noindent
\begin{figure}
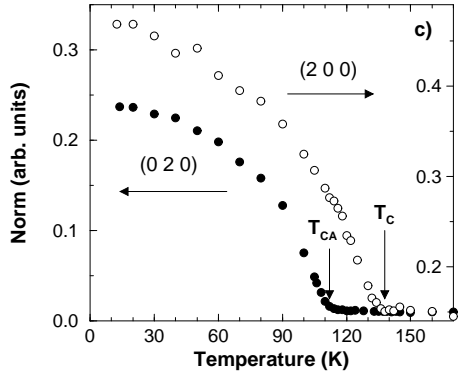

\centerline{\epsfig{file=10_N001.epsi,width=6.cm}}
\centerline{\epsfig{file=10_N002.epsi,width=6.cm}}
\centerline{\epsfig{file=10_N200_020.epsi,width=6.cm}}
\vspace{0.5cm}
\caption{La$_{0.9}$Ca$_{0.1}$MnO$_3$.  Integrated intensities versus temperature 
 ({\bf a}) of (001) ({\bf b}) (002) and  
{\bf c)}  (020) (filled circles) 
and (200) (open circles) Bragg peaks.}
\end{figure}

The concentration dependence of the canting angle $\theta_c$ deduced from observations in the 0$<$x$\le$0.1 range is reported in Fig. 7. A strong 
jump of $\theta_c$ is observed between $x=0.08$ and $0.1$. This abrupt increase of the 
ferromagnetic component is correlated with its rotation in the ({\bf c},{\bf a}) plane. This fast evolution beyond $x=0.08$, departs from the smooth cosine law predicted by de Gennes\cite{de Gennes}. 
\begin{figure}
\centerline{\epsfig{file=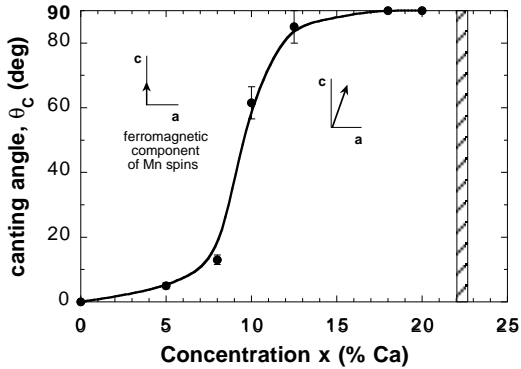,width=7cm}}
\vspace{0.5cm}
\caption{Evolution of the canting angle with doping. The full line is a guide for the eye.}
\end{figure}

{\it At $x=0.125$}, $T_C = 155K$. Below 110K, a very small increase of the (0,0,2l+1) Bragg peaks indicates the occurence of a small antiferromagnetic spin components or a residual canting in this sample. Resistivity measurements performed on this sample, indicate a small decrease just below T$_C$, with an upturn below 110K where this compound becomes insulating\cite{Palstra}. 
Finally, we mention an anomalous increase in the temperature variation of some magnetic Bragg peaks at the temperature called T$_B$$\approx$ 80K (see the (112) Bragg intensity in Fig 5), concomitant with the structural transition temperature T$_{O'O"}$ (cf Fig 2-b). 
\begin{figure}
\centerline{\epsfig{file=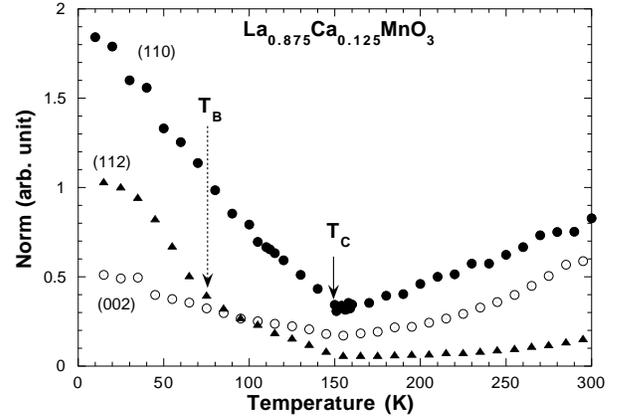,width=8cm}}
\vspace{0.5cm}
\caption{ La$_{0.875}$Ca$_{0.125}$MnO$_3$. Integrated intensities of (110),
(112) and (002) Bragg reflections versus temperature. }
\end{figure}
\noindent
 
{\it At $x=0.2$}, $T_C$ is determined at $185 K$ (Fig. 6). No increase of intensity is detected in the temperature variation of the odd-integer (0,0,2l+1) Bragg peaks, so that this compound is fully ferromagnetic.
The resistivity exhibits a downturn at 
$T_C$ but it still increases at lower temperature\cite{Palstra}.
Our results agree with previous publications, where the metal/insulator transition at T$_C$ has been observed at x$_{Ca}$$\approx$0.22 
\cite{dho,loshkareva}, as indicated by the hatched area in the phase diagram (Fig. 1).

\begin{figure}
\centerline{\epsfig{file=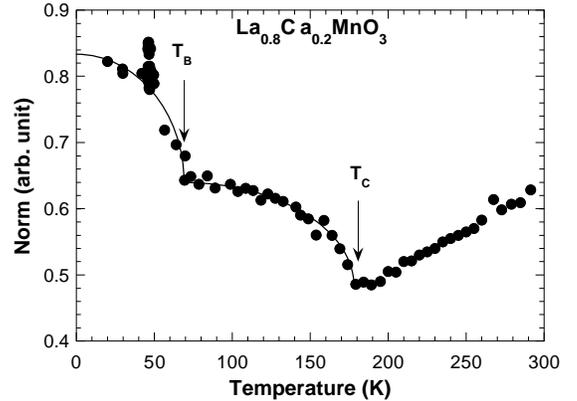,width=8cm}}
\vspace{0.5cm}
\caption{La$_{0.8}$Ca$_{0.2}$MnO$_3$. Integrated intensity of
(110) or (002)  Bragg peaks versus temperature. At 50K, the various points roughly determines two distinct values, observed as a function of time (see the text).}
\end{figure}

 As found for $x=0.125$, an increase of some nuclear and ferromagnetic Bragg peak intensity,  (110) or (002) is also observed at lower temperature, below T$_B$=75K. Actually, its amplitude depends on the rate of cooling.  At 50K, measurements of the Bragg peak (002) have been repeated during several hours. The intensity of the peak and its position are found to fluctuate between two close points (Fig 6). This behaviour is typical of an unstability or a metastability with both magnetic and structural character. It can be attributed to the change in the
 twinned domain structure,  at the structural transition T$_{O'O"}$. 
  Preliminary measurements at $x=0.17$, where the resolution is sufficient to detect the structural transition at T$_B$=T$_{O'O"}$=100K, exhibit a similar anomaly.

{\it In conclusion}, a bell-shape transition line T$_B$ (x) (or T$_{O'O"}$) is defined for x$>$0.1 as reported in Fig 1. It corresponds to a "re-entrant" structural transition where the orthorhombicity is strongly reduced. This line is qualitatively similar to that observed in 
Sr-substituted samples\cite{moritomo} around x=0.125, except that the corresponding temperatures are larger in this latter case. Qualitative differences also appear in the temperature behaviour of some Bragg peaks in the two systems as indicated for the (0,0,2l+1) Bragg peaks. Moreover, the superstructures of the type {\bf Q}= (0,0,l+0.5) which occur below T$_{O'O"}$ in Sr-substituted sample\cite{Yamada} have not been detected in the Ca-subsituted ones. A more extensive structural study of the Ca-doped samples is planned for a complete characterization. \\

{\it 4) Magnetization measurements}\\

Magnetization measurements have been performed as a function of temperature under a 100 Oe applied field and at 10K under large magnetic fields. Small slabs, cut from the four single crystals with x=0.05, 0.08, 0.1 and 0.125 have been studied. The demagnetizing field, similar in all samples, has been minimized by allowing a free alignment of the slab along the field. Fits of magnetization with a Curie-Weiss law 
$\chi=C/(T-\Theta_P)$ show
that $\Theta_P(x)$ strongly increases with x and merges into $T_C(x)$ at $x=0.125$. 
The same variation with x was found in a previous study\cite{matsumoto}.

 Zero-field magnetization can be obtained by extrapolation of the high field measurements to $H=0$, assuming that the volume consists of one domain ( Fig. 7). At $x=0.05$ and 0.08, a quantitative agreement is found with the variation of the canting angle reported in Fig. 4. At $x=0.1$ the magnetization is smaller than expected from the canting angle, which could be explained by an unaccurate determination of the volume due to twinning. 

An interesting 
evolution of the 
coercitive field as a function of $x$ 
is observed, as shown in Fig. 7 and its inset. 
The coercitive field first increases from pure LaMnO$_3$ up to $x=0.1$ and reduces to
 a very small value at $x=0.125$. 
Similar observations were found 
in Sr doped samples 
\cite{paraskevopoulos,popov}. 
 
Such an hysteresis cycle indicates a pinning of the magnetic domains to the lattice, which increases up to x=0.1 and disappears at $x=0.125$. A relation of this evolution with the growth of ferromagnetic clusters, expected to percolate at $x=0.125$ where the antiferromagnetism disappears (cf the next section) is suggested. 
\begin{figure}
\centerline{\epsfig{file=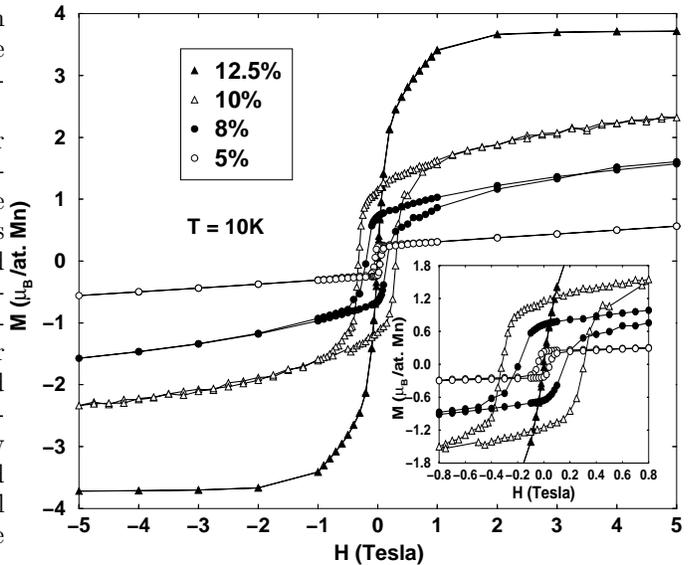,width=10cm}}
\vspace{0.5cm}
\caption{Magnetic field dependence of magnetization at  T=10 K for La$_{1-x}$Ca$_x$MnO$_3$ 
$x=0.05$, $0.08$, $0.1$ and $0.125$. Inset: same measurements displayed for small field values.}
\end{figure}
\noindent
\\
\vspace{0.5cm}

\centerline{\small \bf III. MAGNETIC DIFFUSE SCATTERING}
\vspace{0.5cm}
 
For $x=0.08$, a diffuse scattering, characteristic of short range ferromagnetic correlations, has been already reported\cite{Hennion3}.  In this section, we present the evolution of this diffuse scattering with the Ca concentration in the $0<x \leq 0.125$  range,   
using  two neutron techniques: one, with no energy analysis but a XY multidetector, and the other one, with an energy analysis.\\

{\it 1) Small angle scattering with no energy analysis} \\ 

Small angle neutron scattering (SANS) experiments, using a XY multidetector, have been 
carried out on samples with $x=0, 0.05, 0.08, 0.1$ calcium concentrations at several 
temperatures. The twinned samples were oriented so that the scattering plane is defined by the [110] and [001] directions. Spectra collected in all directions of this scattering plane with the multidetector, indicate that the intensity is nearly isotropic. This allows to check that the various twinned domains have equal volumes. In Fig 8, intensities, corrected for sample transmissions, are compared for several doping rates at $T=10 K$ and along the [110] direction. A strong evolution is observed between  pure LaMnO$_3$ on the one hand, showing a low and nearly q-independent signal, and the doped samples
 on the other hand, characterized 
by a growing q-dependent modulation. This modulation is typical of a characteristic distance between similar ferromagnetic clusters. Interestingly, the same experiment using X-ray 
scattering and a linear multidetector does not show any modulation (see the inset of 
Fig. 8 at $x=0.08$ and T=300K). Since X-ray scattering provides a good contrast between La and Ca ions, we are sure of the absence of any chemical clustering of Ca impurities in the same q range within these samples. 

This modulation lies upon a steeply 
q-decreasing intensity which could be assigned to nuclear scattering (dislocations). In this 
diffraction experiment with no energy analysis, the determination of this parasitic contribution at high temperature cannot be subtracted due to the contamination of the paramagnetic excitations. Only the large q-range, attributed to purely magnetic scattering, can be analysed. A fit with a gaussian function, I(q)$\propto$ exp(-q$^2$R$^2$), indicates an increase 
 of ferromagnetic correlation length from $2R$=14 to 17 and to 19 $\AA$  for x varying from 0.05, 0.08 and 0.1, respectively. 
This is shown in Fig. 9, where ln(I) versus q$^2$ has been reported. We conclude that the ferromagnetic clusters are small, and grow very slowly with x.\\ 

\begin{figure}
\centerline{\epsfig{file=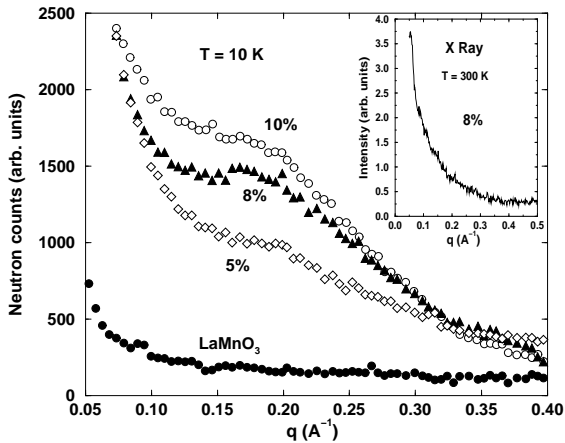,width=7.5cm}}
\vspace{0.5cm}
\caption{Small angle neutron scattering (SANS) spectra measured with a XY multidetector
along [110] direction at $T=10 K$ for La$_{1-x}$Ca$_x$MnO$_3$, $x=0$, 0.05, 0.08 and 0.1. Inset: small angle X-ray measurements at room temperature for x=0.08.}
\end{figure}

\begin{figure}
\centerline{\epsfig{file=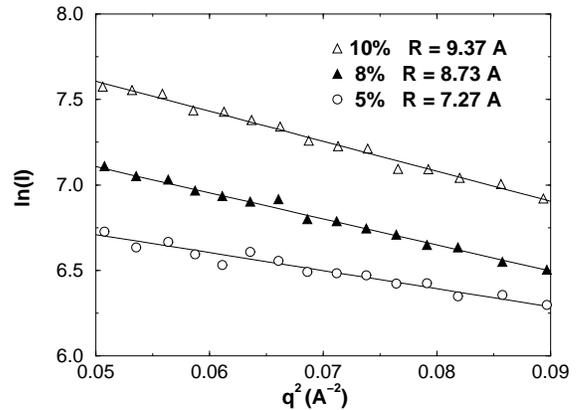,width=7.5cm}}
\vspace{0.5cm}
\caption{Fits of ln(I) versus q$^2$ obtained in the large q range of the SANS spectra, for  La$_{1-x}$Ca$_x$MnO$_3$, $x= 0.05$, 0.08 and 0.1.}
\end{figure}

{\it 2) Small angle scattering with energy analysis}.\\

In order to analyse the  magnetic scattering in the whole q range, elastic diffuse scattering measurements ($\omega=0$) have been performed using a three-axis spectrometer. The $15 \leq T \leq 300 K$ range has been studied along 
several q directions, using $k_i=1.25 \AA^{-1}$. Intensities have been put on an absolute scale, 
using a vanadium sample and a transmission correction. The temperature-independent scattering 
observed above $T_C$ (nuclear contribution) could be subtracted from the low temperature scattering, determining the 
magnetic contribution. This magnetic signal is reported in Fig. 10-a for $x=0.05$, 
0.08, and Fig.10-b for $x=0.1$ and 0.125 at $T=15 K$. Starting from $x=0.05$, the intensity of the modulation rapidly increases
 up to $x=0.08$ with a shift of q, at maximum intensity, towards larger values. Then, at $x=0.1$, the intensity keeps approximately the same value, with a broadening of the modulation indicating additional intensity at smaller q. At $x=0.125$, we no more observe any modulation but the intensity 
is much larger at very small q and a flat q dependence is observed at large q.

A detailed analysis of the $x=0.08$ sample has been previously reported
\cite{Hennion3}. In the picture where the q-modulation results from a characteristic distance between  small "ferromagnetic" clusters, the 
intensity can be expressed as:
\begin{equation}
I(q) \propto | \Delta m^z|^2N_VV^2|F(qR)|^2J(q)
\end{equation}
where $F(qR)$ is the form factor of one cluster, $J(q)$ is a function which characterizes the correlations in the spatial distribution of the clusters (assimilated to a liquid distribution function). The density N$_V$ and the volume $V$, are defined from the q-dependence of $I(q)$ through the $J(q)$ and $F(qR)$ functions. Such a model is based upon two assumptions : (i) The diffuse scattering intensity arises from the  $<S_i^zS_j^z>$ spin correlations, with Oz//{\bf c}. This assumption was actually checked in a previous study on a $x=0.06$ Sr-doped sample, thanks to the absence of twinnning\cite{Hennion2}. The "magnetic" contrast $|\Delta m^z|$ is defined as the difference 
 between the average magnetizations $m^z$ inside and outside the cluster.(ii) The cluster picture is isotropic (spherical shape and 
isotropic cluster distribution in all directions). 
This latter assumption was found to be inaccurate in the twin-free $x=0.06$ Sr doped sample. Instead, a picture of {\it anisotropic} platelets, with a size within the 
ferromagnetic layers about three times larger than perpendicular to them, could be determined. In the present case, equation (1) provides 
a semi-quantitative analysis, with the 3 parameters of the model: the size (2R), the mean distance d$_m$ and a minimal distance of approach d$_{min}$.  The dot-dashed lines in Fig. 10-a are the best fit of the data in this model. The cluster size varies very slowly with x, from $14\AA$ for $x=0.05$, to $17\AA$ for $x=0.08$.
In this approach, the magnetic intensity is mainly proportional to the square of the cluster volume $V$ 
and to the square of the magnetic contrast $\Delta m^z$. Therefore, 
 the increase of the diffuse scattering by a factor $\approx$2 between $x=0.05$ to $x=0.08$, is mainly taken into account by the increase of the cluster size from $14\AA$ to $17\AA$, with a magnetic contrast of
 $\Delta m^z$$\approx$ 0.7$\mu_B$. This is a small contrast, indicating a rather smooth picture. Of course, the assumption of isotropy leads to an overestimation of the volume, or an {\it underestimation} of $\Delta m^z$.  We note that the 
 volume ratio between the clusters observed at $x=0.05$ and at $x=0.08$  
\noindent
roughly compares with the corresponding ratio of the hole concentrations.
\begin{figure}
\centerline{\epsfig{file=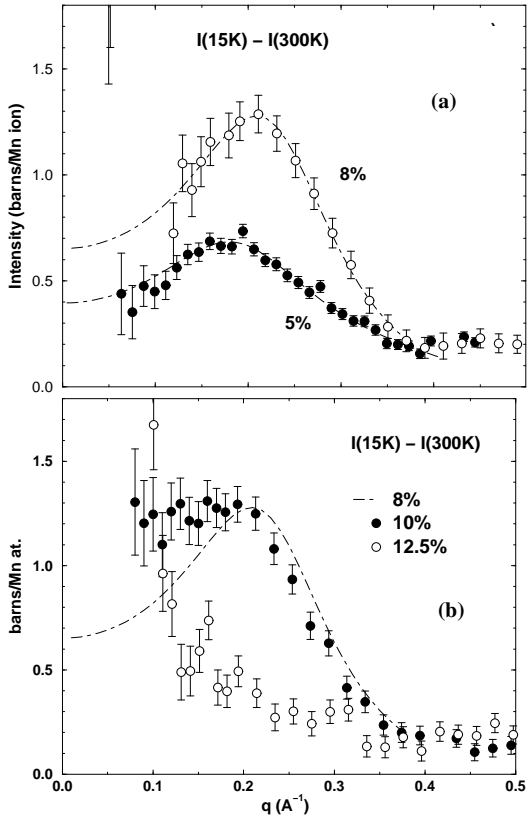,width=7cm}}
\vspace{0.5cm}
\caption{Magnetic scattering observed at small angle in La$_{1-x}$Ca$_x$MnO$_3$ ({\bf a}) $x=0.05$ and $0.08$ and ({\bf b}) $x=0.1$ and $0.15$.}
\end{figure}
\noindent

At $x=0.1$, the evolution is different from that between 0.05 and 0.08.  An additional scattering intensity is observed 
indicating new ferromagnetic correlations on a scale larger than the intercluster distance.  Therefore, the evolution between $x=0.08$ and 0.1, can be characterized in terms of coalescence of clusters, rather
 than in terms of a size variation, as for the evolution between $x=0.05$ and 0.08. 

At $x=0.125$, I(q) shows a strongly increasing intensity at very small q, whereas a flat q-dependence persists at large q. This signal cannot be characterized further. This evolution is likely related to the occurence of the percolation of the ferromagnetic inhomogeneities.

At $x=0.2$, a diffuse scattering of {\it nuclear origin} is observed around (2.5,2,0), that is (.25,2.25,0)$_c$ and equivalent points in reciprocal space in the cubic indexation. This is in perfect agreement with previous observations reported in Ca-doped samples\cite{Dai2}. Since this nuclear diffuse scattering indicating a short-range polaron ordering along the [100] direction cannot be related
to the spin correlations described above (static) and below (dynamic), it is not described further.   
  

The temperature dependence of the magnetic diffuse scattering observed at $x\le 0.1$ has been reported previously\cite{Hennion3} showing that the intensity disappears around the temperature for the magnetic transitionT$_C$.


The existence of ferromagnetic clusters could be the consequence of the polarization by a single mobile hole as previously predicted\cite{Nagaev}. However, in the present case, the comparison between the cluster density and the hole density rather indicates that one ferromagnetic cluster contains several holes. In our first article\cite{Hennion3}, using the assumption of an isotropic model (same distance between clusters whatever the q direction), we have deduced a density of clusters, $\approx$ 1/60 times smaller than that of the holes. The anisotropy discovered in the I(q) function of a twin-free Sr-doped sample, indicates that this value has been overestimated and 
could be rather $\approx$ 1/30. Anyway, this value confirms the picture of a {\it charge segregation} with hole-rich regions embedded in a  Mn$^{3+}$ hole-poor network. 

{\it In conclusion}, we get a picture of  "ferromagnetic " clusters with a diameter of about $16\AA$, in repulsive interaction, which grow very slightly and start to coalesce at $x=0.1$. They are observable as far as the long range antiferromagnetic order exists ($0<x<0.125$). The observation of a characteristic distance between the clusters, as well as the observation of a ferromagnetic Bragg peak, requires that the mean magnetization inside each cluster is parallel to the same direction (here the {\bf c} axis). This
leads to a modulated canted state picture instead of a true phase separation.  

 As shown below, the study of the spin dynamics will corroborate and specify 
this picture. 

\vspace{0.5cm}

\centerline{\small \bf IV. SPIN DYNAMICS}

\vspace{0.5cm}

In section {\bf A}, spin dynamics of the $x=0.1$ compound and in section ({\bf B}) spin dynamics of  $x=0.125$ and $x=0.20$ are reported along the two [001] and [110]  directions of the A-type magnetic structure. In section {\bf C}, the evolution with x of the parameters determined from the spin dynamics is displayed.\\

\centerline{{\bf A) x=0.1: the approach}} 
\centerline{{\bf to the ferromagnetic transition}}
\vspace{0.5cm}

 {\it A-1) Low temperature spin dynamics: comparison with x $<$ 0.1.}\\ 
 {\it A-1-1) The high-energy branch.}\\

As recalled in the introduction, in this low-doped regime, two spin wave dispersion curves are observed, characterized as high-energy and low-energy branches. For the high-energy branch, the assignment of a dispersion to the [001] or the [110] directions is unambiguous, in spite of twinning, because of the two distinct periodicities in the reciprocal space. Exemples of spectra obtained along the [00$\zeta$] direction, which determines the antiferromagnetic coupling, are reported in Fig. 11 up to the antiferromagnetic zone boundary. 
\begin{figure}
\centerline{\epsfig{file=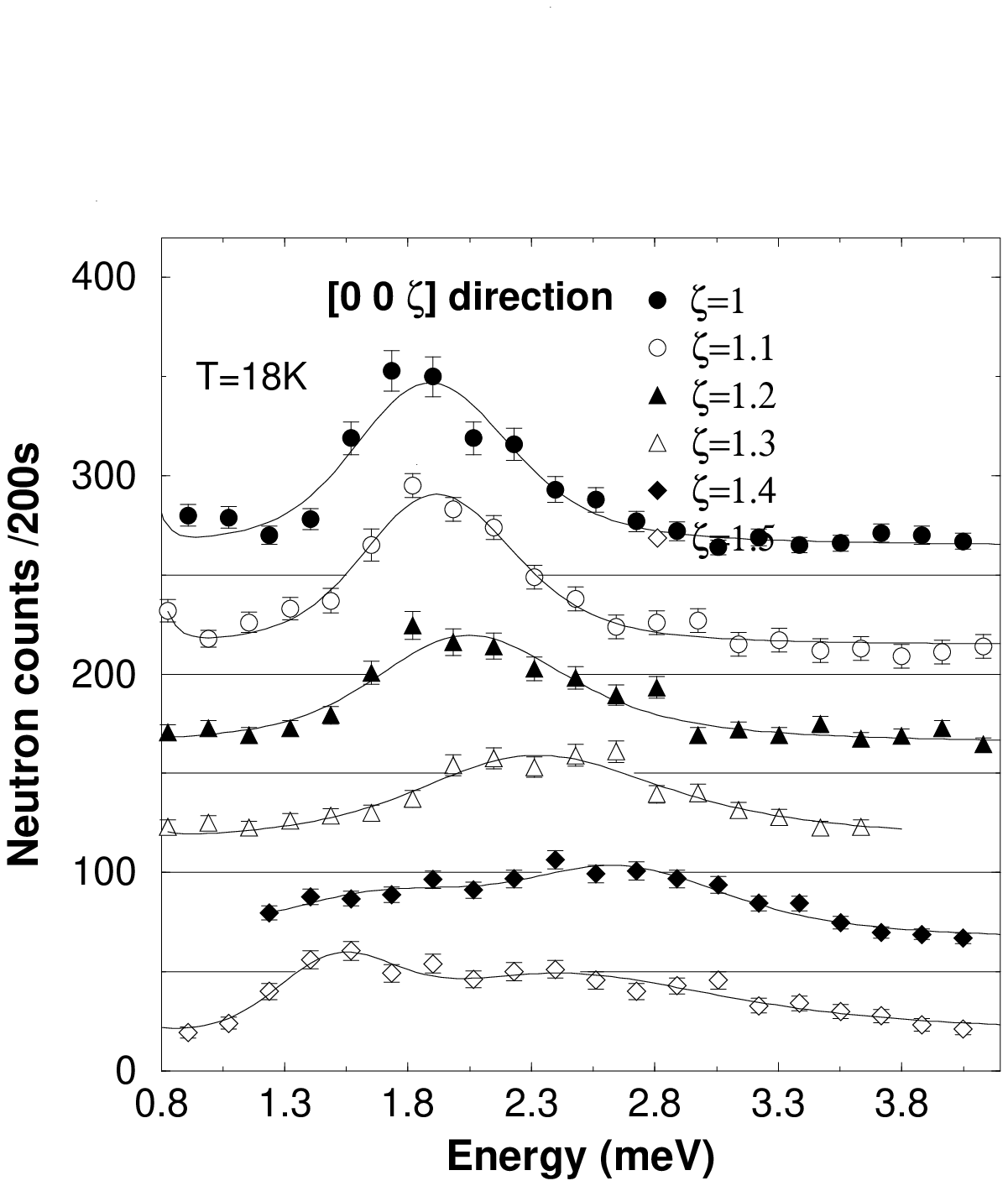,width=7cm}}
\vspace{0.5cm}
\caption{Energy spectra measured in [00$\zeta$] direction up to the antiferromagnetic 
zone boundary, at $T=18K$. Fits correspond to calculated intensities using lorentzian functions convoluted with the spectrometer resolution. }
\end{figure}

 At {\bf Q}=$\tau$=(0,0,1), a large gap value, $\Omega_0$=1.86 meV at $T=18K$, is obtained. By increasing ${\bf q}$ along [001] ({\bf q = Q-$\tau$}), the energy very weakly disperses as the intensity decreases whereas 
the damping, roughly twice larger than in the undoped case, moderately increases. The energy saturates when approaching the zone boundary {\bf Q}=(0,0,1.5) and a lower energy mode belonging
to the low-energy spin wave branch is observed. This high-energy branch has also been measured along the direction of propagation [110] starting from (110), which defines the ferromagnetic coupling.
 There, only spin waves with Q beyond (1.1,1.1,0) 
have a measurable intensity. Therefore, the small q- range along this direction has been studied starting from
the antiferromagnetic Bragg peak (111). The dispersion curve of this high energy branch is reported
 in Fig. 14 by empty triangles along the two symmetry directions.
A fit of the dispersion using a Heisenberg model with four first in-plane neighbor coupling $J_1$, 
ferromagnetic and $J_2$, antiferromagnetic, along [001], and with an effective single ion anisotropy $C$\cite{Moussa1,Hennion1,Moussa2} is also shown by a continuous line in Fig. 14. On the same figure, results obtained for $x<0.1$, and $x=0$, are also reported for comparison.  
The variation of $J_1$ and $J_2$ with doping 
is displayed in Fig. 24.
It reveals a linear variation for the two effective superexchange couplings, showing the weakening of $J_2$ and the 
strengthening of $J_1$. This linear variation J$_2$(x) agrees pretty well with J$_2$ = 0 at $x=0.125$, where the antiferromagnetic Bragg peak (001) disappears.\\

As a remarkable result, the gap $\Omega_0$ of this spin-wave branch keeps the same value 
for all x (see Fig 14 and Fig. 22). This value is also the same, within our experimental accuracy, as that measured in Sr-doped samples\cite{Hennion2,Mukhin}, also reported in Fig. 22. \\

{\it A-1-2) The low-energy branch.}\\

For the case of the low-energy branch, the assignment to the [110] or [001] directions is more difficult, since the periodicities in the q space are the same along the two directions (ferromagnetic character). It can be solved however, as long as the orthorhombicity of the structure is measurable. In q//$\tau$
 experiments, where the Bragg peaks (002) and (110) are a center of symmetry for $\omega(q)$, the assignment is made by performing +q and -q measurements. If the center of symmetry is (002) (resp (110)), the q direction is [001] (resp [110]). As shown in Fig. 12, all the +q and -q magnons lie on a symmetric curve with respect to $q=0$, in the case where the (002) Bragg peak is chosen as q origin. 

\begin{figure}
\centerline{\epsfig{file=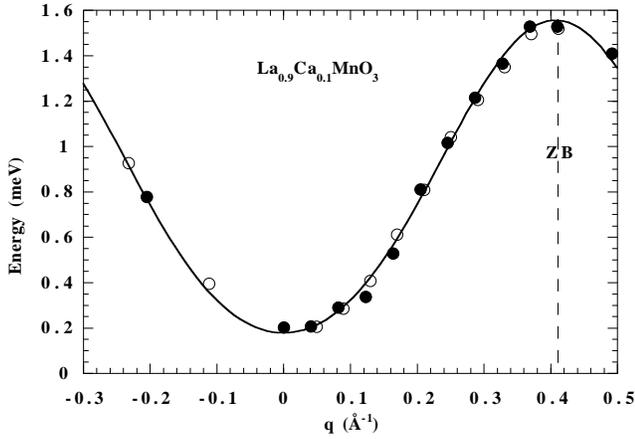,width=8.5cm}}
\vspace{0.5cm}
\caption{Filled and empty circles correspond to energy excitations of the low-energy spin wave branch measured in two perpendicular q directions, attributed to [110] and [001] respectively for one domain. The two corresponding dispersions coincide 
if the q origin is the (002) Bragg peak whatever the domain. 
(+q) and (-q) measurements are symmetric with respect to the (002) Bragg peak (see the text). At this $x=0.1$ concentration, the (002) and (110) Bragg peaks positions are distant from 0.031 \AA$^{-1}$.}
\end{figure}

\begin{figure}
\centerline{\epsfig{file=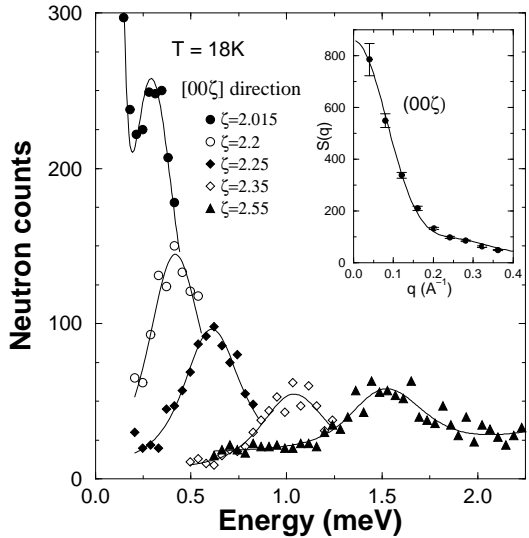,width=7cm}}
\vspace{0.5cm}
\caption{Energy spectra of the low-energy magnetic excitation along [0,0,$\zeta$]. The energy integrated intensity S(q) of this spin-wave branch is shown in the inset.}
\end{figure}

 Exemples of energy spectra are shown in Fig 13, with the energy-integrated intensity S(q) reported in the inset. 
Therefore, the new feature, at this concentration, comes from the fact that spin waves propagating along
[110] cannot be measured. Only propagation along [001] is clearly observed up to the antiferromagnetic zone boundary (cf Fig. 12). 
 The difficulty to measure a dispersion curve along [110], means that the intensity of the corresponding excitations, superimposed to those propagating along  [001], exhibits a very fast decrease with increasing q. Such a strong
anisotropy of the spin wave intensity has already been observed in a Sr-doped sample, where, in absence of twinning,  both directions could be unambiguously identified\cite{Hennion2}. Thanks to a quantitative correspondence between the correlation lengths deduced from the static and dynamic spin correlations, the
 relation with the anisotropic shape of the static clusters has been clearly established. In the same way, we can relate
 the anisotropy of  the spin wave intensity S(q), in this $x=0.1$ sample, to the anisotropy of the shape of the static clusters, hidden here because of twinning (cf III).  Within this picture, the q-profile of the spin wave intensity, S(q), (S(q) is proportionnal to the susceptibility) reported in the inset of Fig. 13, with two distinct q-dependences, can be interpreted as the superposition of the intensities of spin wave propagating along both [110] and [001], characterized by a steep and a slow q-dependence respectively.  The fit of $S(q)$ with a sum of two lorentzian gives two correlation lengths :  $\zeta = 17 \AA$ and $\zeta = 5\AA$. This anisotropy is very similar to the anisotropy of the spin-wave intensity  observed in the twin-free Sr-substituted sample,  which, in this latter case, could be quantitatively compared to the anisotropy of the cluster size\cite{Hennion2}.

 The whole q range of $\omega$(q), along [001] can be fitted by a cosine law, and the small q range, by a $\omega$=$Dq^2$+$\omega_0$ law as in our previous studies. The fitted dispersion is reported in Fig. 14  with experimental data, and compared with the $x=0.05$ and $0.08$ cases. Values of $\omega_0$ and D are reported in Fig. 22 and Fig. 23 respectively showing the decrease of $\omega_0$ and the increase of D with x. 

The comparison of the low-energy spin wave branches reported for x=0.05, 0.08 and 0.1 in Fig 14 reveals that, along [001] where intensity is measurable whatever x, all the dispersion curves bend at the zone boundary 
{\bf Q}=(0,0,1.5),
 or  $q_0$=(0,0,.5) with the {\it same} energy value $\omega$($q_0$) $\approx$ 1.57 meV. 
This observation is the counterpart of the other remarkable observation concerning the x-independent gap $\Omega_0$ of the high energy branch, pointed out above, in the description of the high-energy branch (cf Fig 21).  Both features reveal strongly non-linear effects with x.


The two spin dynamics are tightly coupled. Along [001] they spread into two adjacent energy ranges, defining a small forbidden energy gap (hatched area in Fig. 14). As previously emphasized, they correspond to two types of excitations of a single magnetic ground state, characterized above as a "modulated AF canted " state.

\begin{figure}
\centerline{\epsfig{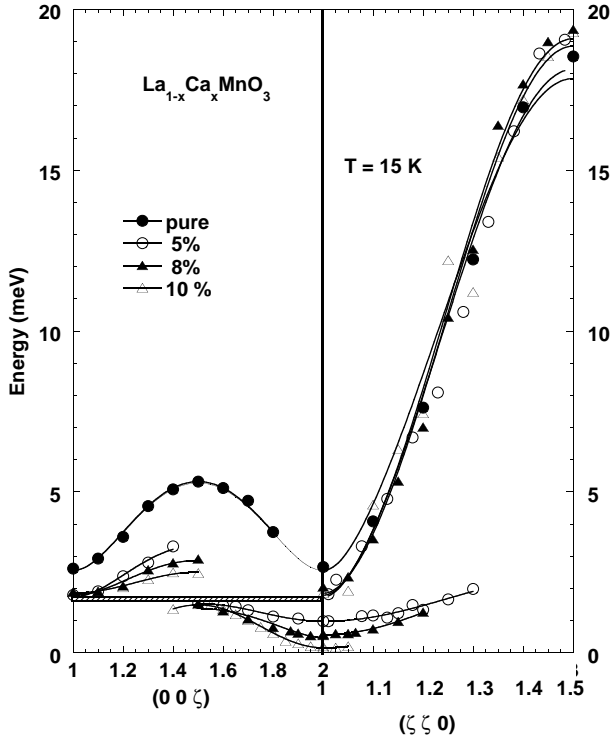}}
\vspace{0.5cm}
\caption{Dispersion curves along [00$\zeta$] from (001) to (002) Bragg peaks (left pannel) and along [$\zeta\zeta$0] from (110) to (1.5,1.5,0) (right pannel) for 
$x=0$, 0.05, 0.08 and 0.1 compounds. At $x=0.1$, along [$\zeta\zeta$0], only the small q excitations of the low energy spin wave branch (empty triangle) have measurable intensity. The solid lines are fits (see the text).}
\end{figure}

\vspace{0.5cm}

\hspace{0.6cm} {\it A-2 Temperature dependence of spin waves}\\
\begin{figure}
\centerline{\epsfig{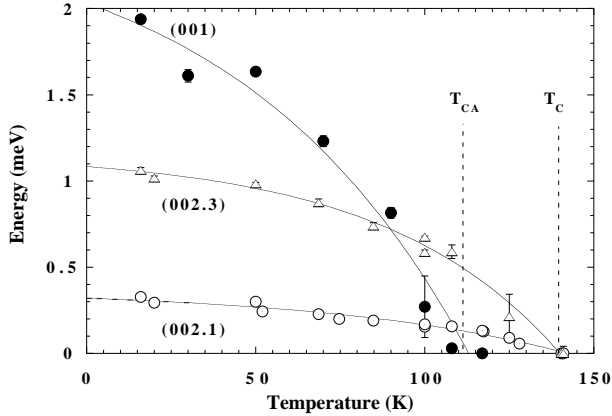}}
\vspace{0.5cm}
\caption{Temperature dependence of some magnetic excitation energies. The full circles correspond to the gap measured at (001). Empty circles correspond to excitations of the low-energy spin-wave branch measured at (002.1) and (002.3). The solid lines are guides for the eye.}
\end{figure}
The phase diagram reported in Fig. 1 has pointed out the existence of a ferromagnetic and insulating phase
between $T_C=138K$ and $T_{CA}=112K$, with a spin direction along {\bf b} at T$_C$ rotating towards {\bf c} below T$_{CA}$, where the antiferromagnetic long range order occurs.  The temperature evolution of the spin dynamics provides the opportunity
to probe more deeply this ferromagnetic state. The temperature evolution of the gap characteristic
 of the high energy branch, measured at the antiferromagnetic Bragg peak $\tau$=(001), is shown by filled circles in Fig. 15. It reveals
a softening of this mode with a complete renormalization at $T_{CA}$. Above $T_{CA}$, we
 still observe AF quasielastic spin fluctuations (energy spectra centered at $\omega$=0).  
 By contrast, the magnetic excitations of the low-energy spin wave branch measured close to the ferromagnetic Bragg peak (002) are still well defined at $T_{CA}$ and are fully renormalized at T$_C$, without any anomaly at T$_{CA}$. This is shown in the same figure, where the temperature evolution of two modes, at q=(002.1) and (002.3), are reported by open circles and triangles, respectively. 
We conclude that the occurence of the long-range ferromagnetic order at T$_C$, is 
driven by the ferromagnetic coupling revealed by the low-energy branch only.
 In section II, we have outlined the similarities of our phase diagram with the phase diagram of de Gennes\cite{de Gennes}. We check here that the small q-range spin dynamics reflect also the mean field theory, whereas inhomogeneous features appear at larger q.

\vspace{0.5cm}
\noindent

\centerline{\bf B) x=0.125 and x=0.2:}

\centerline{\bf The ferromagnetic and insulating phase}
\vspace{0.5cm}

The  concentrations $x=0.125$ and $x=0.2$ correspond to the ferromagnetic and insulating state. Their spin dynamics, at low temperature, is described successively.\\

{\bf 1) x=0.125}.\\

The spin wave dispersions $\omega(q)$ for q along [001] and [110] are reported in the right pannel of Fig. 16,
and in Fig. 17. The assignment of a dispersion curve to the [001] direction of propagation is unambiguous, in spite of twinning, and uses the same arguments as for x=0.1 (cf Fig. 12). The assignment of the other dispersion curve to the [110] direction, is a logical deduction.

The main result is the observation of a single spin wave branch, as expected for a ferromagnetic state. The excitations defining the AF superexchange coupling along [001]
at x$<$0.125 are not observed. This is  what one could expect from the  linear variation of $J_2$(x), where J$_2$ $\approx$0 at this concentration  (Fig. 24)). This single spin wave branch exhibits the same gap as that of the low-energy spin-wave branch in the canted state described above. Therefore, 
a continuity between x $<$ 0.125 and $x=0.125$ is observed for the spin dynamics through the low-energy branch. This observation rules out the idea of a two-dimensional magnetic state at the disappearance of the antiferromagnetism, that could be suggested when missing the low-energy spin-wave branch\cite{Hirota}. 

The dispersion curve differs strongly along [001] and [110], and exhibits some anomaly 
in both directions at {\bf q$_0$} =  ((0,0,1.5) or (1.25,1.25,0)) as described now.

{\it Along [001]}, the dispersion curve is very similar to the low-energy spin-wave branch for x $<$ 0.125. It bends with a vanishing intensity at {\bf q$_0$} =(0,0,0.5), the zone boundary of the antiferromagnetic state, reaching the same x-independent value defined for x $<$ 0.125 (see the horizontal dotted line in Fig. 16). In spite of the disappearance of the antiferromagnetic coupling, the similarity with x$<$0.125 in the dynamic susceptibility implies the persistence of some feature, related to the vicinity of the canted state. In the canted state, we have outlined the tight connexion between the dynamic susceptibility of the low-energy spin wave branch and the static correlations defining the ferromagnetic clusters, attributed to a charge segregation. At x=0.125, this persisting spin dynamic feature may reflect a persisting charge segregation along [001]. 
The gap value and the stiffness constant determined from a fit  $\omega$ = $\omega_0$ + Dq$^2$, are reported in Fig. 22 and 23, respectively. 
   
\begin{figure}
\centerline{\epsfig{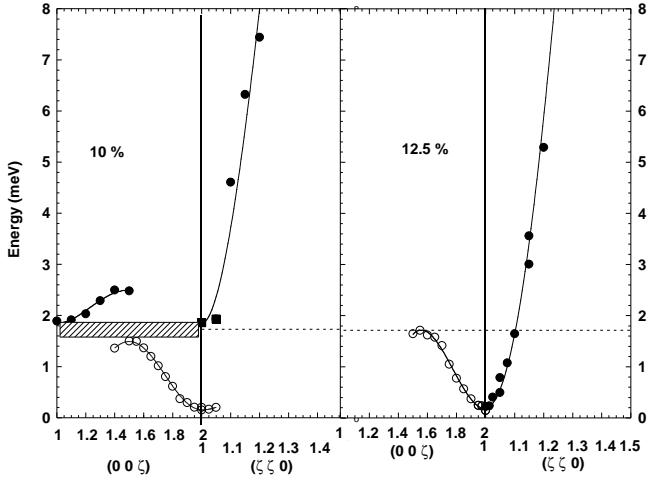}}
\vspace{0.5cm}
\caption{Comparison of the dispersion curves measured along the [00$\zeta$] and [$\zeta\zeta$0]
directions, in the $x=0.1$ (left pannel) and the $x=0.125$ (right pannel). The two squared symbols in the left pannel recall that these excitations have been obtained around (111) instead of (110), where they have no intensity. Continuous 
lines: see the text.}
\end{figure}
\noindent
{\it Along [110]}, $\omega$(q) can be measured up to the ferromagnetic zone boundary (1.5,1.5,0) as shown in Fig. 17, and the dispersion appears anisotropic. The dispersion differs from that observed at x$<$0.125 specially at small q where only the small gap value exists (see comparison with $x=0.1$ in Fig. 17). This strong change at small q suggests that, unlike [001] direction, ferromagnetic platelets have percolated within the ferromagnetic layers, in agreement with the observation of the diffuse scattering (Fig. 10-b). The whole dispersion cannot be fitted by a Heisenberg model with first neighbourg couplings only or a single cosine law, unlike the case of the high-energy branch at lower concentrations. Instead it exhibits an "S" shape, suggesting two different behaviour depending on the lower or larger q ranges considered. In addition, an anomaly appears at half the zone boundary {\bf q$_0$}=(1.25,1.25,0) as seen in Fig. 17. This anomaly can be understood either as a small spitting in the dispersion or as a strong damping of these magnetic excitations, displayed in Fig. 18, where a discontinuity is observed at this {\bf q$_0$}
 value. 

 We mention that a similar anomaly has been reported in La$_{0.85}$Sr$_{.15}$MnO$_3$\cite{vasiliu}, at the same {\bf q$_0$} value, indexed as (0,0,2.5) however, 
and related to the occurence of the new periodicity indicated by the odd-integer superlattice peaks (0,0, 2l+1). 
However,  the present observations are thought to be related to the [110] direction only, instead of [001], so that it cannot be explained by these additional peaks. That will be discussed further with the x=0.2 case.

In spite of the abrupt change in the q$<$$q_0$ low energy-range, $\omega$(q) for q$>$$q_0$ is very comparable to the dispersion of the high-energy branch of the $x=0.1$ sample (see full and empty circles in Fig. 17). An analysis with an Heisenberg model, in terms of SE coupling renormalized by DE ( namely J$_1$) agrees with the linear variation determined for x$<$0.125. It is reported in Fig. 24.

For $q$ $<$ $q_0$, a quadratic law for $\omega(q)$ determines a stiffness constant with a much larger value than along [001], outlining the anisotropic character of the ferromagnetic coupling at this critical concentration.

 \begin{figure}
\centerline{\epsfig{file=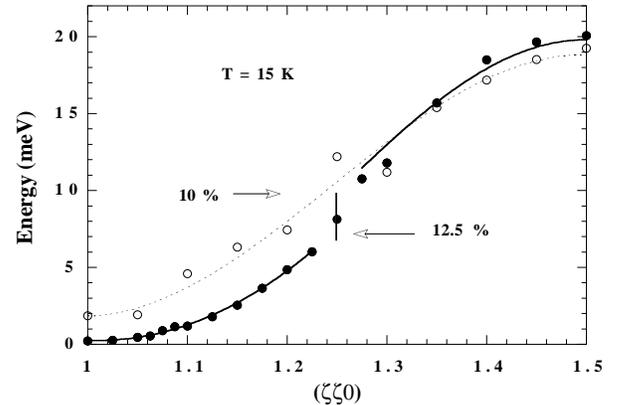,width=8cm}}
\vspace{0.5cm}
\caption{Spin wave dispersion along [110] for x=0.125 (full circles) and x=0.1 (empty circles).}
\end{figure}

 \begin{figure}
\centerline{\epsfig{file=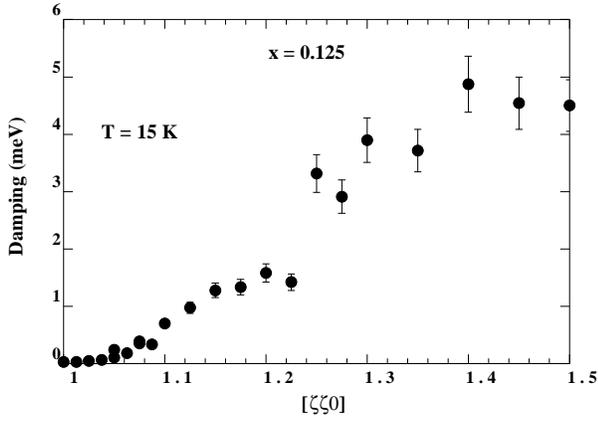,width=8cm}}
\vspace{0.5cm}
\caption{q-dependence of the magnon energy linewidth or damping along [110] for x=0.125.}
\end{figure}

                .

{\bf 2) La$_{0.8}$Ca$_{0.2}$MnO$_3$}

\vspace{0.5cm}

At $x=0.2$, where the transition T$_{OO'}$ occurs very close to $T_C$ (cf the phase diagram of Fig 1), the orthorhombicity is too small for the (110) and the (002) Bragg peaks to be resolved, and odd-integer (0,0,2l+1) Bragg peaks are observed. However, for continuity with the x$<$0.2 concentration range, we still keep the {\it Pbnm} indexation.

The spin wave dispersion obtained at 14K is reported in Fig. 19 for the two superimposed [001] and [110] directions, corresponding to the [100] direction when indexed in the cubic cell. 

\begin{figure}
\centerline{\epsfig{file=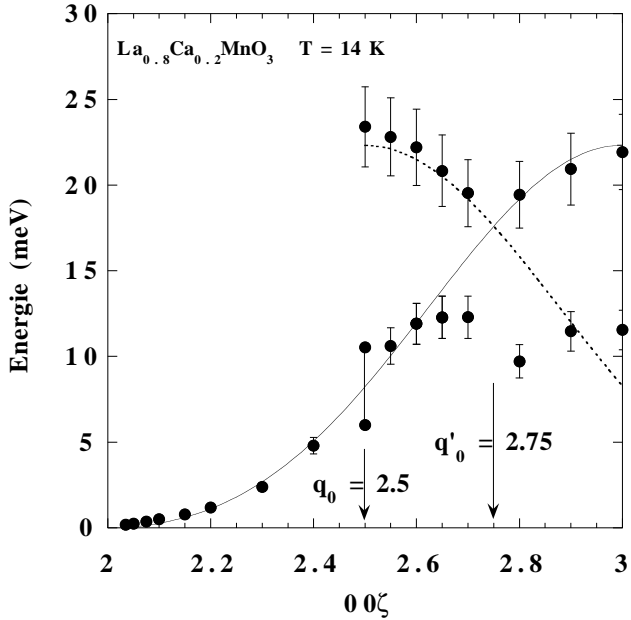,width=8.5cm}}
\vspace{0.5cm}
\caption{Spin-wave dispersions determined at x=0.2 for q along [110] or [001] directions using the orthorhombic unit cell (edge of the small perovskite cube). The continuous line is a guide for the eye. The dotted line shows the anti-crossing behaviour for the two dispersion curves at {\bf q$_0^{'}$}=(0,0,2.75). The splitting into two modes at {\bf q$_0$}=(0,0,2.5) is suggested by the jump in the energy-linewidth at this {\bf q} value, reported in the Fig. 21.}
\end{figure}

At this concentration, we first notice the same anomaly at {\bf q$_0$}=(0,0,2.5) or (1.25,1.25,0) in the q variation of the magnon energy-linewidth (Fig. 21) as in the $x=0.125$ sample. As for x=0.125, the jump of the energy-linewidth at this value suggests a splitting or a folding of the dispersion. However, since at this concentration the [001] and [110] directions are superimposed, such an effect implies the existence of an underlying periodicity for both the [001] and the [110] directions, as provided by [0,0,2l+1) 
and (l+0.5,l+0.5,0) Bragg peaks.

 In addition, new features appears in the larger q range where energy spectra reveal the existence of two types of excitations.  
 At q $\approx$ q$_0$ and beyond, extra excitations are observed at higher energy ($\approx$ 22 meV), forming an optical branch, with a growing intensity as q increases towards the zone boundary. Concomitantly, the intensity of the lower energy excitation decreases.
 Corresponding energy spectra are shown in Fig. 20.  

  Excitations which the largest intensity allow to draw a " main" dispersion curve up to the zone boundary, which, unlike the $x=0.125$ concentration, is identical for [001] and [110] directions. The whole isotropy has been actually checked by spin wave measurements along the two other main symmetry directions, [010] and [101]. 
This "main" spin wave dispersion cannot be fitted with an Heisenberg model using first neighborg coupling only. According to the same analysis as above, for x=0.125, the dispersion within the q$<$q$_0$ range has been fitted using a quadratic law, determining a stiffness constant and a gap values reported in Fig. 23 and Fig. 22 respectively. In the same way as for x=0.125, the energy at the zone boundary may be analysed in terms of a renormalized SE coupling, J$_1$, reported in Fig. 24.

The origin of the additionnal excitations or of the optical branch is very intringuing. The two dispersion curves observed for q$\ge$q$_0$ could be interpreted in terms of an anti-crossing behaviour between the acoustical and the optical branch, related to the folding of the acoustical branch at the reciprocal vector {\bf q$_0^{'}$}=(0,0,2.75) (see the dotted lines in Fig. 19). This could be explained by a periodicity involving {\it four} cubic lattice spacings, such as indicated by (0,0,l+0.5) Bragg peaks.  However, no such peaks have been found in this sample (cf section II). We therefore conclude that, if new atomic correlations occur, they have a short-range or a dynamic nature.

\begin{figure}
\centerline{\epsfig{file=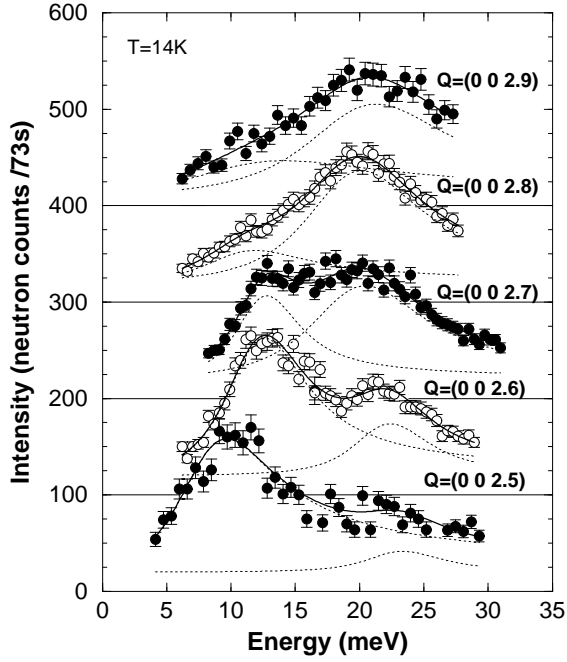,width=7.5cm}}
\vspace{0.5cm}
\caption{Energy spectra measured at x=0.2 along [110] or [001], showing two modes for each q value. The dashed lines correspond to a fit with two lorentzian functions convoluted with the spectrometer resolution. }
\end{figure}

\begin{figure}
\centerline{\epsfig{file=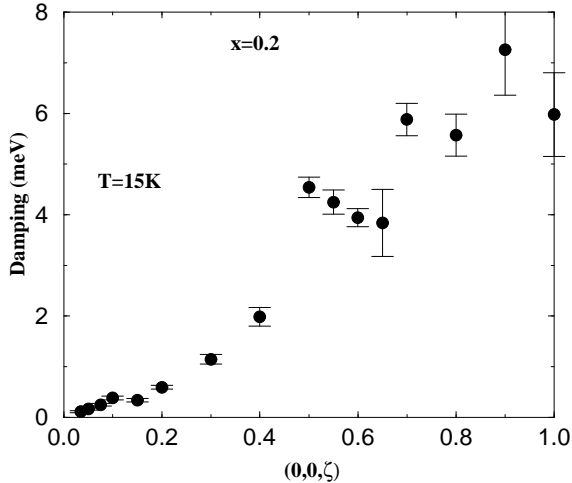,width=7.5cm}}
\vspace{0.5cm}
\caption{Damping versus q of the magnetic excitation measured at x=0.2 along  [001] (or [110]. }
\end{figure}

\centerline{{\bf C)} {\bf Evolution of the dynamic parameters with x.}}
\vspace{0.5cm}

 We report the evolution with x of all the parameters defined from the spin dynamics. \\

{\it a)  Variation with x of the two energy gaps}.\\

 \begin{figure}
\centerline{\epsfig{file=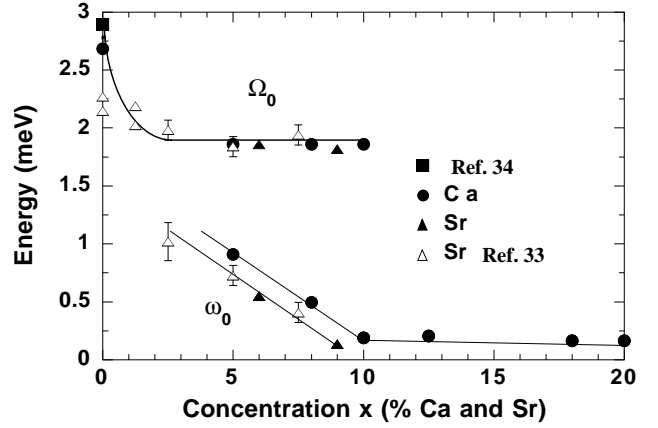,width=8.5cm}}
\vspace{0.5cm}
\caption{Energy gaps corresponding to the two types of spin-wave branches, versus Ca and Sr concentration. Filled symbols: using the neutron technique\cite{Moussa1,Hennion3,Hennion2} at T=15K and\cite{Hirota} at T=8K. Open symbols: using the antiferromagnetic resonance technique\cite{Mukhin} at T$<$20K.}
\end{figure}

In Fig. 22, the  variation with x of the two gaps, the large one $\Omega_0$, associated with the excitations of the A-type magnetic structure and of the small one $\omega_0$, associated with the 
low-energy spin-wave branch, are displayed. On the same figure, measurements on both Ca and 
Sr-doped samples obtained by neutron\cite{Moussa1,Hennion1,Hennion2,Hennion3,Hirota} (filled symbols) and by antiferromagnetic resonance\cite{Mukhin} (empty symbols) are reported. 
 Except for $x=0$ where a small discrepancy is observed between neutron data\cite{Hirota,Moussa1} and AF resonance technique\cite{Mukhin}, the agreement of the data is excellent, whatever the Ca-doped or the Sr-doped samples.
Actually, the low-energy gap reported at $x=0$ by antiferromagnetic resonance\cite{Mukhin}, unresolved by the neutron technique, cannot have the same meaning as $\omega_0$,  related to the low-energy branch, which appears only by doping.

Beyond the very small concentration range (x $\approx$ 0.02) where the large gap $\Omega_0$ decreases fastly, two regimes may be defined, the boundary being around the ferromagnetic transition. 
(i) In the AF canted state (namely x $<$ 0.125 for the Ca substitution and x $<$ 0.1 for the Sr one ), 
$\Omega_0$ is found to be x-independent . The same observation is found for the Sr-doped samples, showing the same constant energy value. As outlined in {\bf IV}A-1, this observation is associated with the existence of a forbidden energy gap (hatched area) separating the energy range of the two spin waves branches along [001] (cf Fig. 14). 
These non-linear features with x, observed both at $q=0$  and at q $\ne$ 0, disagree with the usual picture of a mean-field canted state.  At $q=0$, a monotonous decrease of the large anisotropy gap, $\Omega_0$, associated with the antiferromagnetic structure, is expected\cite{Mukhin}, instead of an x-independent behaviour. Moreover, in such a model, the two spin-wave branches would cross each other along [001] at $q=(0,0,1.5)$,  instead of keeping in separated energy range as shown in Fig. 14.

 As concerning $\omega_0$(x), it fastly decreases up to $x=0.1$, and very slowly decreases beyond, in the ferromagnetic insulating state. 
 Comparison with Sr substitution, indicates that  $\omega_0$(x) is shifted at larger x for Ca than for Sr. in agreement with the experimental fact that Sr induces a stronger ferromagnetic character than Ca.\\
 
{\it b) Evolution of the stiffness constant D with x.}\\

 In Fig. 23, we have reported the variation D(x), obtained for 7 single crystals (x=0.05, 0.08, 0.1, 0.125, 0.17, 0.18, 0.2), deduced from a fit with a quadratic law along [110] or [001] in the small q range only. It shows three regimes.
For x $<$ 0.125, the fit corresponds to the small q-range of the lower energy branch which is {\it isotropic}. Since the corresponding spin wave susceptibility reflects the anisotropic form factor of the ferromagnetic clusters (platelet), this isotropy is surprising. Likely, the stiffness constant D may reflect a mean coupling, which would involve the magnetic coupling both inside the clusters and between the clusters through the canted magnetic state of the hole-poor medium.

 At $x=0.125$, where a single spin wave branch is observed, the dispersion differs along [001] and [110]. The step-increase observed along [110], corresponds to the collapse between the two spin wave branches into a single one, whereas the low value of D along [001] is in perfect continuity with the evolution of the low-energy spin-wave branch. As discussed previously, the absence of change in the long distance ferromagnetic coupling along [001] and the abrupt increase of this coupling along [110], agree with a picture of percolation of the platelets occuring in the ferromagnetic layers only.  This implies a persisting spin disorder from one layer to the neighbour one.
At x=0.175 and beyond, far from the canted state limit, D appears as isotropic. 
Interestingly, a continuity of D(x) could be suggested, when fitting the whole variation by a quadratic function of x.

\begin{figure}
\centerline{\epsfig{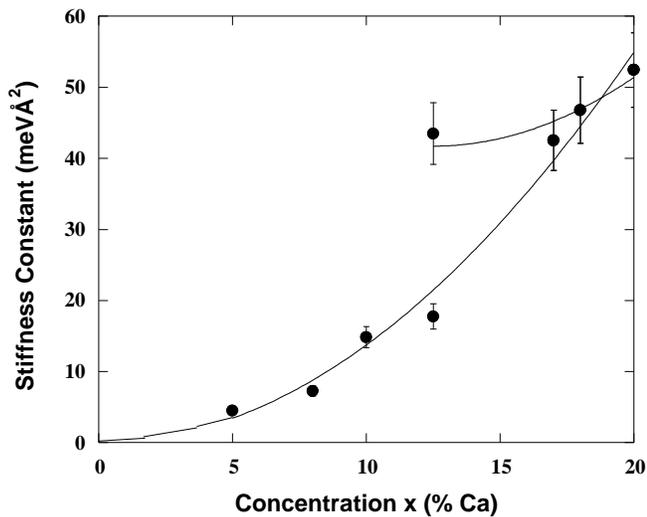}}
\vspace{0.5cm}
\caption{Stiffness constant determined from the low-energy spin-wave branch versus Ca concentration. The values at $x=0.17$ and  0.18 are deduced from a preliminary study. }
\end{figure}

{\it c) Evolution of the "effective" superexchange constants with x.}\\

In Fig. 24, the ferromagnetic coupling J$_1$ and the antiferromagnetic coupling J$_2$ are reported as a function of x. For x $<$ 0.125, J$_1$ and J$_2$ are deduced from the high-energy branch along [110] and [001] directions respectively, whereas for x$>0.1$, J$_1$ is deduced from the energy around the zone boundary of the Brillouin zone of the ferromagnetic state. A linear variation is observed with x. It yields J$_2$$\approx$ 0 close to $x=0.125$, in agreement with the observation of a single ferromagnetic spin-wave branch
 at this concentration (cf above). The renormalization of SE coupling by hole doping agrees very well with the theoretical work of Feiner and Oles\cite{Feiner}. However, since it does not account for the canted state, this model leads to a 2-d behavior at some concentration (J$_2$=0) and to J$_2$$>$0 beyond. By contrast, we show the continuous increase of the 3-d ferromagnetic coupling with x through the ferromagnetic low-energy branch.

\begin{figure}
\centerline{\epsfig{file=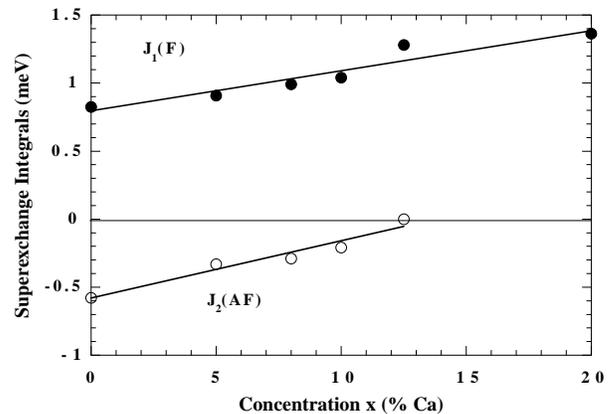,width=8cm}}
\vspace{0.5cm}
\caption{Ferromagnetic coupling J$_1$ and antiferromagnetic coupling J$_2$, associated with SE type of coupling (see the text)
as a function of Ca concentration.}
\end{figure}

\vspace{0.5cm}

\centerline{\small \bf VI. CONCLUSION}

\vspace{0.5cm}

The three studied concentrations $x=0.1$, 0.125 and 0.2, described in this paper, characterize three distinct steps in the evolution of the magnetic ground state towards the insulator-metal transition.
    
{\it The  x=0.1 sample} exhibits the same original properties reported in the very low-doping regime ($x=0.05$ and 0.08), i. e. an additionnal spin wave branch of ferromagnetic character and a static diffuse scattering, characteristic of ferromagnetic inhomogeneities in interaction. The new observation here, is the
 strongly anisotropic q dependence of the intensity of the low-energy spin wave branch. This may be related to the two-dimensional (2-d) character of the ferromagnetic clusters, or platelets. This feature, readily observed in the diffuse scattering of a free-twin Sr doped sample\cite{Hennion2}, is difficult to observe in the static diffuse scattering reported here because of the twinning. These 2-d ferromagnetic clusters are embedded in an average canted state. The proposed analysis of the diffuse scattering which yields a density of clusters much smaller than the density of holes, suggests a {\it charge segregation}. This {\it charge segregation} picture differs from the {\it phase separation} picture , with antiferromagnetic and ferromagnetic states, predicted by the theory\cite{Khomskii,Kagan,Nagaev,Dagotto,Yunoki,Arovas}. A description in terms of modulated canted state is the most appropriate. We outline, however, the unusual q-dependence of the spin-waves,  which appears to be specific of this canted state induced by charge doping.

{\it The $x=0.125$ sample} is the limit where the antiferromagnetism of the A-type structure disappears. This ferromagnetic state results from two evolutions with x, occuring concomitantly: a linear decrease of the antiferromagnetic coupling of SE type, J$_2$ where J$_2$ $\approx$0 at $x=0.125$ (homogeneous process) and, as revealed by the diffuse scattering, 
the growth of the ferromagnetic clusters, suggesting a kind of percolation of the hole-rich clusters at this concentration (inhomogeneous process). This magnetic percolation does not coincide with the percolation for transport properties (metallic state), since an insulating behaviour is still observed  at low temperature. At this concentration, the spin dynamics consists of a single spin wave branch, as expected for a ferromagnet, but with an anisotropic dispersion. From the inhomogeneous picture derived at smaller concentration, this anisotropy could be explained by the percolation of the platelets occuring in the basal plane only (2-d percolation). Whatever, in the DE picture, this anisotropy implies some persisting spin misalignment between one layer and its neighbour ones. 

 In addition, an anomaly is observed along [110] at q$_0$ in the spin wave dispersion, corresponding to half the zone boundary. 
 
{\it In the x=0.2 sample} (nominal concentration), the spin wave dispersion appears isotropic. The anomaly observed at q$_0$, as for x=0.125, is now assigned to the two superimposed [001] and [110] directions. For the direction [001], such an anomaly could be related to the new structural periodicity indicated by the odd-integer [0,0,2l+1] peaks, which involves two cubic lattice spacings. However, no explanation can be found for the [110] direction where such peaks are not observed. Therefore, we cannot exclude another origin, occuring both along [110] and [001]. Interestingly, a recent spin wave calulation\cite{Mancini}, has shown the importance of the antiferromagnetic coupling between the localized $t_{2g}$ spins of the Mn for the calculation of the spin-waves. This antiferromagnetism could explain the underlying periodicity. Whether this theoretical work, which describes the metallic state, may have any relevance very close to the metallic sate, remains to be clarified. 

The newest feature at this concentration is however the observation of additional excitations or of an optical branch at larger q and larger energies. This observation suggests an anti-crossing behaviour or a folding of the acoustical branch at the reciprocal vector of the type (0,0,2.75). Such a feature could be explained by an additional periodicity involving 4 lattice spacings, giving rise to superstructures
at {\bf Q}=(0,0,l+0.5). However, as emphasized in section II, these peaks have not been observed in this sample. We mention that these peaks have been actually observed in Sr-substituted samples with x=0.125\cite{Yamada} below the low-temperature structural transition called T$_{O'O"}$ or T$_B$ in the present study.  A subsequent work using X-ray resonant scattering\cite{Endoh}, describes this new structural distorsion in terms of a new orbital ordering. Due to the strong similarity usually found between Sr and Ca substitution, a structural distorsion could also exist here, but on a short-range scale or with a dynamic character, so that no Bragg peak intensity could be detected.
The splitting or the anti-crossing feature also suggests that the magnon excitations are interacting with other excitations, magnetic or not.  A physical process of magnon-phonon coupling has been suggested by Furukawa\cite{Furukawa} in these systems, whereas other authors\cite{Khaliulin,Oles} have proposed a coupling between magnon and orbital excitations.  Although proposed to explain the anomalous softening of magnon observed in the metallic state\cite{Dai}, such a coupling can be expected also in the present case where at low temperature the "reentrant" structural transition indicates a disordered orbital state.  Only a neutron study with polarised neutrons will unambiguously define the true nature, purely magnetic or not, of these additional excitations. Finally, the evolution of the spin dynamics with temperature, specially below T$_B$ where the re-entrant structural transition is observed, will be also important for a deeper understanding of this complex magnetic state. It will be reported in a forthcoming paper.

\vspace{0.5cm}

{\bf Aknowledgements}
\vspace{0.5cm}

The authors have very indebted to L. Noirez from Laboratoire Leon Brillouin  for his help during the small angle neutron experiments and A. Wildes from Institut Laue Langevin (Grenoble). They further aknowledge M. Viret, B. B. van Aken, T. T. M. Palstra, A. Moreo, D. I. Khomskii, E. L. Nagaev and A. M. Oles for stimulating discussions.

\end{document}